\documentclass[a4paper,10pt]{article}
\usepackage[a4paper,text={13cm,20cm},centering,scale=0.7]{geometry}
\usepackage{amsfonts,amsmath,amsthm,amssymb,bm,graphicx}
\title{Riemannian geometry without the hypotheses of homogeneity and symmetry}
\author{Paolo Maraner\footnote{School~of~Economics~and~Management, Free University of Bozen-Bolzano, Universit\"atsplatz-Piazzetta dell'Universit\`a~1, I-39100 Bolzano-Bozen, Italy, email: pmaraner@unibz.it}}
\date{}
\begin{document}
\parindent=5pt
\maketitle
\begin{abstract}
A generalisation of Riemannian geometry is considered, based exclusively on the minimal assumptions that the line element $ds$ is a regular function of position and direction and that the distance of every point from itself is equal to zero. Besides the Riemannian line element, also Riemann's residual hypotheses of homogeneity and symmetry are dropped. Surprisingly, the infinitesimal Pythagorean distance formula reemerges, without the need of being postulated, as a first approximation to \emph{almost every} geometry that is invariant with respect to direction reversal. More in general, the first approximation to \emph{almost every} geometry is a one-parameter family of homogeneous Riemann-Randers line elements, naturally providing the geometrical framework for a unified theory of the classical electromagnetic and gravitational fileds. Geometry naturally accounts for the hierarchy between electromagnetic and gravitational interactions, their different attractive/repulsive nature, electric charge, CPT symmetry, Maxwell and Einstein equations. Within this framework higher order terms could describe new spacetime degrees of freedom of possible large-scale relevance.
\end{abstract}

\section{Introduction}
Geometry is central to our understanding of the world. Viewed as something real or
as a convention to describe reality, it has evolved with our thinking and will continue to do so \cite{Jammer1954}. After the centuries needed to conceive what we  today call the Euclidean space, the advent of calculus and  non-Euclidean geometries on the one side, and the investigation of electromagnetic phenomena on the other, have produced a radical transformation of our concept of space. The turning point is marked by Riemann's trial lecture \emph{On the Hypotheses That Lie at The Foundations of Geometry} \cite{Riemann,Pesic2007}.  In just over half a century, Riemann's vision has become the backbone of the general theory of relativity and therefore of our understanding of spacetime. This has stimulated an enormous interest in differential geometry, aiming to reconcile Einstein's theory of gravity with the other fundamental interactions and with the quantum world. 

The purpose of this work is far less ambitious. 
I am going to investigate a differential geometry based solely on the minimal assumption that the distance of every point from itself is equal to zero. Thus, besides renouncing to the 
infinitesimal Pythagorean distance formula as in Finsler geometry \cite{Rund2012,BaoChernShen2012}, I will also drop Riemann's residual hypotheses of homogeneity and symmetry. Surprisingly, this leads to a logically simple construction, very close in spirit and techniques to standard Riemannian geometry.
In particular, I will show that the infinitesimal Pythagorean distance formula reemerges, without the need of being postulated, as a first approximation to \emph{almost every}  geometry that is invariant with respect to direction reversal. More in general, \emph{almost every} geometry turns out to be approximated by a one-parameter family of homogeneous Riemann-Randers metrics  \cite{Randers1941}. The parameter is related to a geometric conservation law. Homogeneity is recovered for each value of the parameter, but not as a whole. The connection and the curvature associated to a generic line element  emerge from the identification of geodesic equations with auto parallel equations, leading to a univocal generalisation of the Christoffel-Levi-Civita connection and of the Riemannian curvature tensor.

If, following the spirit of general relativity we apply this geometry to spacetime, 
we effortlessly  find ourselves with what could well be the unified theory of the classical electromagnetic and gravitational fields, long sought by Weyl, Kaluza, Eddington, Einstein, Schr\"{o}dinger and many others \cite{Goenner2004,Goenner2014,Tonnelat1966}. Compared to other attempts, the identification of geometrical objects with physical quantities is logical and completely straightforward, like the imposition of the possible field equations. Moreover, some fundamental yet puzzling facts, such as the hierarchy between electromagnetic and gravitational interactions, their different attractive/repulsive nature and CPT invariance, naturally emerge from the geometrical framework. Whether this geometry will bring any progress in our comprehension of the classical spacetime it is too early to say. 
Nonetheless, the conceptual simplicity and economy that it offers in the formulation of both the electromagnetic and gravitational theories seems to me undeniable.\\

In section \ref{hypotheses} I will review Riemann's hypotheses on geometry together with  their revisions, from the indefiniteness of the line element in pseudo-Riemanninan geometry, to line elements that are arbitrary homogeneous functions of direction in Finsler geometry. A differential geometry based on a line element only constrained by the assumption that the distance of every point from itself equals zero is presented in section \ref{relax}. Section \ref{spacetime}
describes the possible applications to spacetime. Two appendices respectively discuss the geometrical interpretation of the gravitational field in general relativity and offer an elementary introduction to nonlinear connections.

\section{Riemann's hypotheses on geometry}\label{hypotheses}
In his celebrated 1854 trial lecture \emph{On the Hypotheses That Lie at The Foundations of Geometry}, Riemann introduces a general analytic approach to geometry by extending Gauss's work on curved surfaces to ``$n$-fold extended manifolds"\footnote{If not otherwise stated, all the quotations throughout this paper are from Riemann's \emph{Ueber die Hypothesen, welche der Geometrie zu Grunde liegen} \cite{Riemann} in the English translation by M. Spivack with slight revisions by P. Pesich \cite{Pesic2007}.}. These are defined as spaces parameterized by $n$ independent coordinates $x^\mu$, $\mu=1,2, ...,n$. 
The investigation of metric relations of such spaces is approached by measuring the length of lines, so that the distance between two points is implicitly defined as the length of the shortest line joining them, a geodesic line. Primarily, Riemann assumes that as ``measurement requires independence of quantity from position [...] the length of lines is independent of their configuration, so that every line can be measured by every other", a concept that Einstein will better rephrase by stating that ``if two line segments are found to be equal at one time and at some place, they are equal always and everywhere"\cite{Einstein1921}. Next, to set up a mathematical expression for the length of lines, Riemann treats the problem ``only under certain restrictions" by assuming
\begin{description}
  \item[\emph{Differentiability}:] The lines whose length is to be determined are differentiable, so that the problem reduces ``to setting up a general expression for the line element $ds$'' connecting two infinitesimally adjacent points  $x=\left(x^1, x^2, ..., x^n\right)$ and $x+dx=\left(x^1+dx^1, x^2+dx^2, ...,x^n+dx^n\right)$, ``an expression which will involve the quantities $x$ and the quantities $dx$.'' Therefore,  $ds=f(x,dx)$.
  \item[\emph{Homogeneity}:] If all increments $dx^\mu$ are  increased by the same ratio, $ds$ must also be increased by the same ratio. Equivalently, $ds$ must be  ``an arbitrary homogeneous function of the first degree in the quantities $dx$''. That is to say, $f(x,k\, dx)=k f(x,dx)$ with $k>0$.
  \item[\emph{Symmetry}:] The line element ``remains the same when all quantities $dx$ change sign.'' Equivalently, $ds$ is an even function of the increments $dx^\mu$. 
In symbolic notation, $f(x,-dx)=f(x,dx)$.
\end{description}
At this point, in order to find the explicit form for $ds$, Riemann seeks  ``an expression for the $(n-1)$-fold extended manifold which are everywhere equidistant from the origin of the line element, i.e.\ [...] a continuous function which distinguishes them from one another. This must either decrease or increase in all directions from the origin;" he assumes that ``it increases in all directions and therefore has a minimum at the origin. Then if  its first and second differential quotients are finite, the first order must vanish and the second order cannot be negative;" he eventually assumes that ``it is always positive." From this he deduces that
\begin{description}
 \item[\emph{Quadratic restriction}:] The line element $ds$ equals ``the square root of an everywhere positive homogeneous function of the second degree in the quantities $dx$, in which the coefficients are continuous functions of the quantities $x$.''\footnote{The expression \emph{quadratic restriction} to describe Riemann's choice of the infinitesimal Pythagorean distance formula 
for the line element was introduced by Chern in \cite{Chern1996}.}  
\end{description}
Namely, the infinitesimal expression of the Pythagorean distance formula in arbitrary coordinates. In modern notation
\begin{equation}
\label{Rds}
ds=\sqrt{g_{\mu\nu}dx^{\mu}dx^{\nu}},
\end{equation}
where $g_{\mu\nu}(x)$ are regular functions of the coordinates and the summation over repeated indices is understood. Clearly, the fourth hypothesis implies the previous three but is not implied by them. In particular, the assumption that the ``second differential quotient[...]'' is finite, does not imply that it is nonvanishing. From this Riemann realised that the minimum can also be produced by an appropriate root of a  biquadratic or higher-order expression. In fact, he notes that ``[t]he next simplest case would perhaps include the manifolds in which the line element can be expressed as the fourth root of a differential expression of the fourth degree."\\

Remarkably, Riemann makes clear that these hypotheses ``are not logically necessary," and ``one can therefore investigate their likelihood, which is certainly very great within the bounds of observation, and afterwards decide on the legitimacy of extending them beyond the bounds of observation". As is well known, the `certainty within the bounds of observation' soon called for a revision of Riemann's fourth hypothesis. This is implicitly contained in Minkowski's geometrical reformulation of Einstein's theory of special relativity and was later made explicit in the general theory.  The unification of space and time into a single four dimensional continuum that ``sprang from the soil of experimental physics"\cite{Minkowski1908} required the revision of the \emph{quadratic restriction} allowing the indefiniteness of the quadratic form appearing in $ds$ or, equivalently, allowing pseudo-Euclidean geometries besides the Euclidean one on infinitesimal scales.

Another call for revision came some thirty years later from Randers's observation that ``the most characteristic property of the physical world is the uni-direction of time-like intervals" so that  ``the perfect symmetry in any direction for any coordinate interval does not seem quite appropriate for the application to physical space-time"\cite{Randers1941}. In other words, the simultaneous parity and time reversal symmetry PT is not a fundamental symmetry of nature, becoming such only if accompanied by the simultaneous transformation of charge conjugation C. Accordingly, it would be preferable to have something like CPT symmetry embodied in the fundamental texture of spacetime geometry, rather than PT \emph{symmetry}.

Riemannian geometry without the \emph{quadratic restriction}, and also with or without the \emph{symmetry} hypothesis, has been developed since the 1920s by Finsler, Bernwald, Buselman, Cartan, Rund, Chern and many others \cite{Rund2012,BaoChernShen2012}. For historical reasons it is known as Finsler geometry. The construction, that requires ``no essential new ideas"\cite{Chern1996}, is far from being logically simple, leading to a structure supporting at least three different curvatures and five torsion tensors. 
Moreover, ``while Riemannian geometry can be handled, elegantly and efficiently, by tensor analysis [...] Finsler geometry [...] needs more than one space [...] on which tensor analysis does not fit well"\cite{Chern1996}. Starting from the 1930s attempts have been made to apply Finsler geometry to spacetime, both in the effort to extend the theory of relativity or to unify gravity with electromagnetism \cite{Goenner2004,Goenner2014,Tonnelat1966,Rutz1993,Asanov2012,Pfeifer&Wohlfarth2012}. However, in spite of the extreme richness in structure, it ``seems that this particular generalisation of Riemannian geometry is not able to lead to a correct implementation of the electromagnetic field"\cite{Stephenson1957}, nor to provide a logically attractive extension of general relativity. Riemann's observation that the ``investigation of this general class [...] require[s] no essential different principles [...] and throw[s] proportionally little new light on the study of space" seems to be fully justified, after all. 

At present, \emph{homogeneity}  is  considered a fundamental technical requirement in Riemannian geometry. In fact, it heuristically justifies the chain of identities
\begin{equation*}
\int ds=\int{\textstyle\frac{1}{d\tau}} f(x,dx)d\tau=\int
f\left(x(\tau),{\textstyle\frac{dx}{d\tau}(\tau)}\right)d\tau
\end{equation*} 
that allows to explicitly evaluate the line integral of $ds$  and, when extrapolated to $k=0$,  implies that the distance of every point from itself equals zero. This latter really seems  to be an unrenounceable requirement for a distance function. In fact, in the abstract theory of metric spaces it is one of the axioms that a distance function must satisfy. However, nothing like homogeneity is generally required for a distance function. It is therefore tempting to consider the possibility of relaxing Riemann's \emph{homogeneity} hypothesis to the weaker assumption of
\begin{description}
  \item[\emph{Indiscernibility}:] If all increments $dx^{\mu}$ are vanishing, $ds$ must also vanish. That is to say, $f(x,0)=0$.\footnote{That the distance of every point from itself is zero, is generally assumed together with its inverse statement (if the distance between two points is zero then the two points coincide) under the name of \emph{identity of indiscernibles}. By analogy, I will call the direct statement alone \emph{indiscernibility} as the line element does not allows to perceive a point as two distinct entities. Given the existence of hyperbolic geometry the inverse statement does not  fit as an axiom for geometry. }
\end{description}
To the best of my knowledge, no one so far has considered Riemannian geometry without the hypotheses of \emph{homogeneity} and \emph{symmetry}.

\section{Relaxing homogeneity and symmetry}\label{relax}
Hereafter I will consider Riemann-type geometries uniquely based on the hypotheses of \emph{differentiability} and \emph{inderniscibility}. For the sake of simplicity I will assume differentiability in its strongest sense, i.e.\ analyticity.
\subsection*{Line element}
The keystone that supports the construction of Riemann-type geometries without the hypothesis of \emph{homogeneity},  is the fact that the line element can again be taken in the general form  
\begin{equation}
\label{ds}
ds=f\left(x(\tau),{\textstyle\frac{dx}{d\tau}(\tau)}\right)d\tau
\end{equation} 
with $f(x,\dot{x})$ an arbitrary analytic function of the coordinates $x$ and of the directions $\dot{x}=\frac{dx}{d\tau}$ and with the \emph{proviso} that \emph{the parameter $\tau$ is implicitly determined by the geodesic equations}. Shortly, I will show that the geodesic equations do actually determine the geodesic parameter $\tau$. In complete generality I will proceed by series-expanding $ds$ about the point $\dot{x}=0$. The line element is then rewritten as
\begin{equation}
\label{tds}
ds=\left(g_\mu\dot{x}^\mu+\frac{1}{2}g_{\mu\nu}\dot{x}^\mu\dot{x}^\nu+
\frac{1}{3!}g_{\mu\nu\xi}\dot{x}^\mu\dot{x}^\nu\dot{x}^\xi+...+
\frac{1}{k!}g_{\mu_1...\mu_k}\dot{x}^{\mu_1}...\ \dot{x}^{\mu_k}+...\right)d\tau
\end{equation}
where the zero order term is missing because of the hypothesis of \emph{indiscernibility}  and the symmetric tensors $g_\mu(x)=\frac{\partial f}{\partial \dot{x}^\mu}(x,0)$, $g_{\mu\nu}(x)=\frac{\partial^2 f}{\partial \dot{x}^\mu\partial \dot{x}^\nu}(x,0)$, ..., $g_{\mu_1...\mu_k}(x)=\frac{\partial^k f}{\partial \dot{x}^{\mu_1}...\partial \dot{x}^{\mu_k}}(x,0)$ ..., are regular functions of the coordinates. It is important to emphasize that these tensors are not independent geometrical objects but only a different way of encoding all the information contained in the line element $ds$. The variation of the line integral of $ds$ produces now the geodesic equations in terms of the symmetric tensors $g_{\mu_1...\mu_k}(x)$  and of their derivatives. However, we do  not need to write them down explicitly in order to see that they do determine the geodesic parameter. As $ds$ does not depend explicitly on $\tau$, from the analogy with analytical mechanics we can immediately conclude that the geodesic equations admit the first integral
\begin{equation}
\label{fi} 
g_{\mu\nu}\dot{x}^\mu\dot{x}^\nu+ \frac{2}{3}g_{\mu\nu\xi}\dot{x}^\mu\dot{x}^\nu\dot{x}^\xi+...+ \frac{2}{k(k-2)!}g_{\mu_1...\mu_k}\dot{x}^{\mu_1}...\ \dot{x}^{\mu_k}+...=2E,\end{equation}
with $E$ an arbitrary constant. This identity implicitly determines $d\tau$ in terms of the differentials $g_{\mu_1...\mu_k}dx^{\mu_1}...dx^{\mu_k}$ and the constant $E$. Consequently, it determines $\tau$ up to an irrelevant additive constant. \\

For example, if $g_{\mu\nu}dx^\mu dx^\nu>0$ and $E>0$, as is always the case in standard Riemannian geometry, the series-inversion of equation (\ref{fi}) yields \begin{eqnarray}\label{dtau}
d\tau=\frac{1}{2\varepsilon}\sqrt{g_{\mu\nu}dx^\mu dx^\nu}+
\frac{g_{\mu_1\mu_2\mu_3}dx^{\mu_1}dx^{\mu_2}dx^{\mu_3}}{3g_{\mu\nu}dx^\mu dx^\nu}+\hskip118pt
&&  \nonumber \\[5pt]
+\varepsilon \left(\frac{ g_{\mu_1\mu_2\mu_3\mu_4}dx^{\mu_1}dx^{\mu_2}dx^{\mu_3}dx^{\mu_4}}{4\sqrt{\left(g_{\mu\nu}dx^\mu dx^\nu\right)^3}}-\frac{\left(g_{\mu_1\mu_2\mu_3}dx^{\mu_1}dx^{\mu_2}dx^{\mu_3}\right)^2}{3\sqrt{\left(g_{\mu\nu}dx^\mu dx^\nu\right)^5}}\right)
+...&& 
\end{eqnarray}
where, for simplicity, I have set $\varepsilon=\sqrt{E/2}$. The substitution of 
(\ref{dtau}) into (\ref{tds}) yields the explicit form of the line element in terms of  ``the quantities $x$ and the quantities  $dx$" as sought by Riemann
\begin{eqnarray}
\label{gdsExp}
\frac{ds}{\varepsilon}=\frac{1}{\varepsilon}g_\mu dx^\mu+
\sqrt{g_{\mu\nu}dx^\mu dx^\nu}+
\hskip193pt
&&  \nonumber \\[5pt]
-\varepsilon^2 
\left(\frac{g_{\mu_1\mu_2\mu_3\mu_4}dx^{\mu_1}dx^{\mu_2}dx^{\mu_3}dx^{\mu_4}}{6\sqrt{\left(g_{\mu\nu}dx^\mu dx^\nu\right)^3}}-\frac{2\left(g_{\mu_1\mu_2\mu_3}dx^{\mu_1}dx^{\mu_2}dx^{\mu_3}\right)^2}{9\sqrt{\left(g_{\mu\nu}dx^\mu dx^\nu\right)^5}}\right)
+...&& 
\end{eqnarray}
Quite surprisingly, the most general line element for  Riemann-type geometries without the hypothesis of \emph{homogeneity} turns out to be a one-parameter family of homogeneous line elements. 

For geometries that are symmetric under direction reversal, where  $g_{\mu}$ as well as all odd-order tensors $g_{\mu_1\mu_2... \mu_{2k-1}}$ vanishes identically, the (rescaled) line element reduces in first approximation to the standard Riemannian line element $ds=\sqrt{g_{\mu\nu}dx^\mu dx^\nu}$. If terms of order $\varepsilon^2$ and higher can be neglected, the geometry turns out to be homogeneous because the line element can always be rescaled by a constant factor. This shows the extraordinary generality of Riemann's assumption. In this limit (\ref{Rds}) can be derived from the sole hypotheses of \emph{differentiability}, \emph{indiscernibility} and \emph{symmetry}. In particular, the Pythagorean proposition emerges, \emph{without the need of being postulated}, as the fundamental building block of geometry on infinitesimal scales. 
Clearly, it is still possible to construct non-Pythagorean geometries\footnote{If we repeat the series inversion and the substitution of $d\tau$ in $ds$ under the assumption of a vanishing $g_{\mu\nu}$, we obtain that ``the [rescaled] line element can be expressed as the fourth root of a differential expression of fourth degree'' plus correction of order $\varepsilon$ and higher,
 $${\textstyle\frac{3}{\varepsilon^{3/2}}}ds=\sqrt[4]{g_{\mu_1\mu_2\mu_3\mu_4}dx^{\mu_1}dx^{\mu_2}dx^{\mu_3}dx^{\mu_4}}+O(\varepsilon).$$
This was lucidly foreseen by Riemann's deep insight.} where $g_{\mu\nu}\equiv0$. However, the number of these geometries equals the number of analytic functions with vanishing Hessian at $\dot{x}=0$. Consequently, non-Pythagorean geometries only represent a subset of zero measure of the set of all possible Riemann-type geometries.

For generic geometries, the leading contribution to the line element is given by the one form $g_\mu dx^\mu$, which is however insufficient to specify a geometry. Consequently,  geometry is described in first approximation by a one-parameter family of homogeneous Riemann-Randers metrics \cite{Randers1941}
\begin{equation}
\label{RRds}
ds=\frac{1}{\varepsilon}g_\mu dx^\mu+\sqrt{g_{\mu\nu}dx^\mu dx^\nu}.
\end{equation}
As $\varepsilon$ can no longer be eliminated by a constant rescaling,  the geometry as a whole is no longer homogeneous. On the other hand, while non-symmetrical by construction, the (rescaled) line element remains the same to all orders, if the transformation $dx^\mu\to-dx^\mu$ is accompanied by the sign reversal of the parameter $\varepsilon\to-\varepsilon$.\\

The general case is not substantially different but requires a careful discussion of different cases in order to express the line element $ds$ as a function of ``the quantities $x$ and the quantities $dx$". On the other hand, this process is completely unnecessary. The geodesic equations can be more simply and conveniently derived by the direct variation of the line integral of (\ref{tds}).

\subsection*{Geodesic equations}
 The Euler-Lagrange equations for the lines giving a stationary value to the line integral of  (\ref{tds}), the geodesic equations of this geometry, are obtained as
\begin{eqnarray}\label{ge}
\left[g_{\kappa\lambda}
+g_{\kappa\lambda\mu}\dot{x}^{\mu}+...
+\frac{1}{(k-2)!}g_{\kappa\lambda\mu_1... \mu_{k-2}}\dot{x}^{\mu_1}...\ \dot{x}^{\mu_{k-2}}+...\ \right]\ddot{x}^{\lambda}+
\hskip20pt
&&  \nonumber \\[5pt]
+\Gamma_{\kappa\mu}\dot{x}^{\mu}
+\Gamma_{\kappa\mu\nu}\dot{x}^{\mu}\dot{x}^{\nu} 
+\Gamma_{\kappa\mu\nu\xi}\dot{x}^{\mu}\dot{x}^{\nu}\dot{x}^{\xi}
+... 
+\Gamma_{\kappa\mu_1...\mu_k}\dot{x}^{\mu_1}...\ \dot{x}^{\mu_k}+ ...
=0, \hskip-20pt 
\end{eqnarray}
where dots indicate differentiation with respect to the parameter $\tau$ and the symbols $\Gamma$ are defined as
\begin{eqnarray}\label{Gammas}
&&\Gamma_{\kappa\mu} = \partial_\mu g_\kappa-\partial_\kappa g_\mu \nonumber 
\\[5pt]
&&\Gamma_{\kappa\mu\nu}  = \frac{1}{2}\left(\partial_\mu g_{\kappa\nu}
+\partial_\nu g_{\mu\kappa}-\partial_\kappa g_{\mu\nu}\right)  \nonumber\\[5pt]
&&\Gamma_{\kappa\mu\nu\xi}  = \frac{1}{3!}\left(\partial_\mu g_{\kappa\nu\xi}
+\partial_\nu g_{\mu\kappa\xi}+\partial_\xi g_{\mu\nu\kappa}
-\partial_\kappa g_{\mu\nu\xi}\right)  \\[5pt]
&& \hskip10pt ...\nonumber\\[5pt]
&&\Gamma_{\kappa\mu_1...\mu_k}=\frac{1}{k!}
\left(\partial_{\mu_1}g_{\kappa\mu_2...\mu_{k-1}}+ ...
+\partial_{\mu_n}g_{\mu_1...\mu_{k-1}\kappa} -
\partial_\kappa g_{\mu_1...\mu_k}
\right)\nonumber\\[5pt]
&& \hskip10pt ...\nonumber
\end{eqnarray}

Up to an overall sign, the $\Gamma_{\kappa\mu}$ are related to $g_\mu$
in the same way as an electromagnetic field is related to its vector potential. 
Also, the $\Gamma_{\kappa\mu\nu}$ correspond to the Christoffel symbols of the first kind associated to the metric $g_{\mu\nu}$. In the light of (\ref{Gammas}) these two relationships appear now as the special cases $k=1$ and $k=2$ of a more general rule that links a symmetric tensor $g_{\mu_1...\mu_k}$ of order $k$, to  the symbols $\Gamma_{\kappa\mu_1...\mu_k}$.

 Apart from $\Gamma_{\kappa\mu}$,  that transforms like a tensor, all $\Gamma$ symbols display non-covariant transformation rules, somehow analogues to that of the Christoffel symbols. Nonetheless, equations (\ref{ge}) can be given an explicit covariant form by introducing a single symmetric linear connection ${C^\kappa}_{\mu\nu}$. This can possibly, but not necessarily, be chosen as the Christoffel-Levi-Civita connection ${\Gamma^\kappa}_{\mu\nu}=g^{\kappa\lambda}{\Gamma}_{\lambda\mu\nu}$ associated to $g_{\mu\nu}$. In order to maintain full symmetry in the formalism, for the moment I will assume  that  ${C^\kappa}_{\mu\nu}\neq {\Gamma^\kappa}_{\mu\nu}$. Nothing of what follows depends in any way on the choice of ${C^\kappa}_{\mu\nu}$ (see Appendix A). 
 
 By adding and subtracting identical terms on the left hand side of (\ref{ge}),  the geodesic  equations can be rewritten in the form
\begin{eqnarray}\label{cge}
\left[g_{\kappa\lambda}
+g_{\kappa\lambda\mu}\dot{x}^{\mu}+...
+\frac{1}{(k-2)!}g_{\kappa\lambda\mu_1... \mu_{k-2}}\dot{x}^{\mu_1}...\ \dot{x}^{\mu_{k-2}}+...\ \right]\left(\ddot{x}^{\lambda}+{C^\lambda}_{\nu\xi}\dot{x}^\nu\dot{x}^\xi\right)+\hskip2pt
&&  \nonumber \\[5pt]
+{F}_{\kappa\mu}\dot{x}^{\mu}
+{F}_{\kappa\mu\nu} \dot{x}^{\mu}\dot{x}^{\nu}
+{F}_{\kappa\mu\nu\xi} \dot{x}^{\mu}\dot{x}^{\nu}\dot{x}^{\xi}
+ ... 
+{F}_{\kappa\mu_1...\mu_k}\dot{x}^{\mu_1}...\ \dot{x}^{\mu_k}+ ...
=0,  && 
\end{eqnarray}
where the second derivatives $\ddot{x}^{\lambda}$ have been made covariant by the addition of appropriate terms and the non-covariant $\Gamma$ symbols have been replaced by the covariant tensor fields
\begin{eqnarray}
&&{F}_{\kappa\mu}=\Gamma_{\kappa\mu} = \nabla_\mu g_\kappa-\nabla_\kappa g_\mu \nonumber\\[5pt]
&&{F}_{\kappa\mu\nu}  
=\Gamma_{\kappa\mu\nu}-g_{\kappa\lambda}C^\lambda_{\mu\nu}
= \frac{1}{2}\left(\nabla_\mu g_{\kappa\nu}
+\nabla_\nu g_{\mu\kappa}-\nabla_\kappa g_{\mu\nu}\right)\nonumber\\[5pt]
&&{F}_{\kappa\mu\nu\xi}  
=\Gamma_{\kappa\mu\nu\xi}-g_{\kappa\lambda(\mu}C^\lambda_{\nu\xi)}
= \frac{1}{3!}\left(\nabla_\mu g_{\kappa\nu\xi}
+\nabla_\nu g_{\mu\kappa\xi}+\nabla_\xi g_{\mu\nu\kappa}
-\nabla_\kappa g_{\mu\nu\xi}\right) 
\nonumber\\[5pt]
&& \hskip10pt ...\\[5pt]
&&{F}_{\kappa\mu_1...\mu_k}=\Gamma_{\kappa\mu_1...\mu_k}-\frac{1}{(k-2)!}g_{\kappa\lambda(\mu_1... \mu_{k-2}}C^\lambda_{\mu_{k-1}\mu_k)}=\nonumber\\[5pt]
&&\hskip40pt =\frac{1}{k!}
\left(\nabla_{\mu_1}g_{\kappa\mu_2...\mu_{k-1}}+ ...
+\nabla_{\mu_n}g_{\mu_1...\mu_{k-1}\kappa} -
\nabla_\kappa g_{\mu_1...\mu_k}
\right)\nonumber\\[5pt]
&& \hskip10pt ...\nonumber
\end{eqnarray}
where $\nabla_\kappa$ is the covariant derivative associated to the linear connection ${C^\kappa}_{\mu\nu}$. We note that the tensor fields $F$ are simply obtained from  (\ref{Gammas}) by replacing the partial derivatives $\partial_\kappa$ with the covariant derivatives $\nabla_\kappa$. 

The second order symbols $\Gamma_{\kappa\mu}=F_{\kappa\mu}$ are unaffected by this transformation. The Christoffel symbols of the first kind $\Gamma_{\kappa\mu\nu}$ associated to $g_{\mu\nu}$, are instead transformed into the analogues of the covariant gravitational field ${F}_{\kappa\mu\nu}$ of what can be described as the Poicar\'e-Rosen formalism for general relativity (see Appendix A). 
All other higher order Christoffel-like symbols are transformed into the covariant fields ${F}_{\kappa\mu_1...\mu_k}$.
It is interesting to note that even in this general approach to Riemannian geometry, the quadratic term in the series expansion of $ds$ maintains a unique status. The corresponding field ${F}_{\kappa\mu\nu}$, is the only one that produces local effects indistinguishable from a coordinate transformation. Correspondingly, ${F}_{\kappa\mu\nu}$ can always be set identically equal to zero by choosing  ${C^\kappa}_{\mu\nu}$ equal to the Christoffel-Levi-Civita connection associated to $g_{\mu\nu}$. Since this considerably simplifies the formalism, making it analogous to that of standard Riemannian geometry and general relativity, from now on I will choose  ${C^\kappa}_{\mu\nu}= {\Gamma^\kappa}_{\mu\nu}$. With this choice the geodesic equations are rewritten as 
\begin{eqnarray}\label{sge}
\left[g_{\kappa\lambda}
+g_{\kappa\lambda\mu}\dot{x}^{\mu}+...
+\frac{1}{(k-2)!}g_{\kappa\lambda\mu_1... \mu_{k-2}}\dot{x}^{\mu_1}...\ \dot{x}^{\mu_{k-2}}+...\ \right]\left(\ddot{x}^{\lambda}+{\Gamma^\lambda}_{\nu\xi}\dot{x}^\nu\dot{x}^\xi\right)+\hskip2pt
&&  \nonumber \\[5pt]
+{F}_{\kappa\mu}\dot{x}^{\mu}
+{F}_{\kappa\mu\nu\xi} \dot{x}^{\mu}\dot{x}^{\nu}\dot{x}^{\xi}
+ ... 
+{F}_{\kappa\mu_1...\mu_k}\dot{x}^{\mu_1}...\ \dot{x}^{\mu_k}+ ...
=0,  && 
\end{eqnarray}
where ${F}_{\kappa\mu\nu}$ no longer appears and $g_{\mu\nu}$ plays a privileged role with respect to the other fields.
Nonetheless, it is important to stress that all the $g_{\mu_1...\mu_k}$ take part in shaping the metric properties of space, such as the length of lines and the distance between points. In principle all fields can be treated on an equal footing.\\

Let us now consider the effect of an arbitrary reparameterization $\tau\to\zeta(\tau)$ of the geodesic equations. A direct computation shows that  (\ref{sge}) transforms into
\begin{eqnarray}\label{zrcge}
\left[g_{\kappa\lambda}
+...
+\frac{{\dot{\zeta}}^{k-2}}{(k-2)!}g_{\kappa\lambda\mu_1... \mu_{k-2}}\dot{x}^{\mu_1}...\ \dot{x}^{\mu_{k-2}}+...\ \right]\left(\ddot{x}^{\lambda}+{\Gamma^\lambda}_{\nu\xi}\dot{x}^\nu\dot{x}^\xi+\frac{\ddot{\zeta}}{{\dot{\zeta}}^{2}}\dot{x}^\lambda\right)+\hskip5pt
&&  \nonumber \\[5pt]
+\dot{\zeta}^{-1}{F}_{\kappa\mu}\dot{x}^{\mu}
+\dot{\zeta}{F}_{\kappa\mu\nu\xi} \dot{x}^{\mu}\dot{x}^{\nu}\dot{x}^{\xi}
+ ... 
+\dot{\zeta}^{k-2}{F}_{\kappa\mu_1...\mu_k}\dot{x}^{\mu_1}...\ \dot{x}^{\mu_k}+ ...
=0. && 
\end{eqnarray}
On the one side, this confirms that the geodesic equations determine the geodesic parameter $\tau$ up to an arbitrary additive constant. In fact, the simpler form (\ref{sge}) can hold only for a particular parameter $\tau$, distinguished up to the linear transformation $\tau\to\tau+\tau_0$ with constant $\tau_0$. This transformation is clearly responsible for the conservation of (\ref{fi}).
On the other side, we learn that the geodesic equations are unaffected by a  simultaneous constant rescaling of the parameter $\tau$ and of the fields  $g_{\mu_1...\mu_k}$. In fact, the equations (\ref{sge}) are invariant under the combined  transformation
\begin{equation}
\label{rescaling}
\tau\to\lambda\tau+\tau_0
\hskip10pt\text{and}\hskip10pt
g_{\mu_1...\mu_k}\to\lambda^{k-2}g_{\mu_1...\mu_k},
\end{equation}
with $\lambda$ a nonzero constant.
Correspondingly, the line element (\ref{tds}) and the first integral (\ref{fi}) 
are rescaled by an irrelevant factor $\frac{1}{\lambda^2}$, 
\begin{equation}
\label{rescaling1}
ds\to\frac{1}{\lambda^2}ds
\hskip10pt\text{and}\hskip10pt
E\to\frac{1}{\lambda^2}E.
\end{equation}
In investigating non-null geodesics, we are therefore free to directly solve the equations (\ref{sge}), whose solutions are labeled by $E$, or to stipulate a standard gauge on $\tau$ and solve the rescaled equations (\ref{zrcge}) for a fixed  value of first integral (\ref{fi}). This last option is customarily made in standard Riemannian geometry and in general relativity, where the geodesic parameter is conventionally identified with the arc length $s$, so that $d\tau=\sqrt{g_{\mu\nu}dx^\mu dx^\nu}$ and $g_{\mu\nu}\dot{x}^\mu \dot{x}^\nu=1$.
The corresponding gauge in Riemannian geometry without the hypotheses of \emph{homogeneity} and \emph{symmetry} corresponds to a rescaling by $\lambda=\varepsilon=\text{sgn}(E)\sqrt{|E|/2}$.  This produces the geodesic equations in the form
\begin{eqnarray}\label{rcge}
\left[g_{\kappa\lambda}
+...
+\frac{\varepsilon^{k-2}}{(k-2)!}g_{\kappa\lambda\mu_1... \mu_{k-2}}\dot{x}^{\mu_1}...\ \dot{x}^{\mu_{k-2}}+...\ \right]\left(\ddot{x}^{\lambda}+{\Gamma^\lambda}_{\nu\xi}\dot{x}^\nu\dot{x}^\xi\right)+\hskip10pt
&&  \nonumber \\[5pt]
+\frac{1}{\varepsilon}{F}_{\kappa\mu}\dot{x}^{\mu}
+\varepsilon{F}_{\kappa\mu\nu\xi} \dot{x}^{\mu}\dot{x}^{\nu}\dot{x}^{\xi}
+ ... 
+\varepsilon^{k-2}{F}_{\kappa\mu_1...\mu_k}\dot{x}^{\mu_1}...\ \dot{x}^{\mu_k}+ ...
=0, && 
\end{eqnarray}
with  $d\tau=\pm\sqrt{|g_{\mu\nu}dx^\mu dx^\nu|}+O(\varepsilon)$ and the first integral (\ref{fi}) equal to $\pm1$. \\

Equations (\ref{sge}) and equivalent ones, generalise the geodesic equations of standard (pseudo-)Riemannian geometry, to which they reduce when all $g_{\mu_1...\mu_k}$ except $g_{\mu\nu}$ vanish identically. By assuming the invertibility of the second order tensor
\begin{equation}
\label{ }
\bm{g}_{\mu\nu}
= g_{\mu\nu}
+g_{\mu\nu\xi}\dot{x}^{\xi}+...
+\frac{1}{(k-2)!}g_{\mu\nu\xi_1... \xi_{k-2}}\dot{x}^{\xi_1}...\ \dot{x}^{\xi_{k-2}}+...\ ,
\end{equation}
these equations can be resolved with respect to the second derivatives. The local existence and uniqueness theorem for their solutions follows then from  standard ODE-theory.

\subsection*{Connection and curvature}
The standard form of the geodesic equations is also a convenient starting point to identify the connection and the curvature associated to the line element (\ref{tds}). By series-inverting $\bm{g}_{\mu\nu}$ we obtain the symmetric contravariant tensor $\bm{g}^{\mu\nu}$ fulfilling the identity $\bm{g}^{\mu\kappa}\bm{g}_{\kappa\nu}=\delta^\mu_\nu$. The first few terms are  given by
\begin{equation}
\bm{g}^{\mu\nu}=g^{\mu\nu}
-{g^{\mu\nu}}_\xi\dot{x}^\xi
+\left[
{g^{\mu}}_{\lambda(\xi}{g^{\nu\lambda}}_{o)}
-\frac{1}{2}{g^{\mu\nu}}_{\xi o}
\right]\dot{x}^{\xi}\dot{x}^{o}+ ...\ ,
\end{equation}
where 
round brackets indicates symmetrisation  and indices are raised by means of the inverse $g^{\mu\nu}$ of $g_{\mu\nu}$, $g^{\mu\kappa}g_{\kappa\nu}=\delta^\mu_\nu$. For example, ${g^{\mu\nu}}_\xi=
g^{\mu\rho}g^{\nu\sigma}g_{\rho\sigma\xi}$. By multiplying (\ref{sge}) by $\bm{g}^{\iota\kappa}$, contracting on $\kappa$ and renaming indices, we  obtain the geodesic equations in the standard form 
\begin{equation}
\label{gesf}
\ddot{x}^\kappa+\bm{\gamma}^\kappa_\mu\dot{x}^\mu=0
\end{equation}
with $\bm{\gamma}^\kappa_\mu\left(x,\dot{x}\right)$ expressed as a series in $\dot{x}$ as
\begin{equation}
\label{nlco}
\bm{\gamma}^\kappa_\mu={\gamma}^\kappa_\mu
+{\gamma}^\kappa_{\mu\nu}\dot{x}^\nu
+{\gamma}^\kappa_{\mu\nu\xi}\dot{x}^{\nu}\dot{x}^{\xi}
+...+{\gamma}^\kappa_{\mu\xi_1...\xi_n}\dot{x}^{\xi_1}...\ \dot{x}^{\xi_n}+... 
\end{equation}
The first few coefficients are given by 
\begin{eqnarray}
&&  {\gamma}^\kappa_\mu={F^{\kappa}}_\mu
\nonumber\\[5pt]
&&  {\gamma}^\kappa_{\mu\nu}=
{\Gamma^\kappa}_{\mu\nu}
-{g^{\kappa}}_{\lambda(\mu} {{F}^{\lambda}}_{\nu)}
\nonumber\\[5pt]
&&  {\gamma}^\kappa_{\mu\nu\xi}={F}^{\kappa}_{\ \mu\nu\xi}
-{g^{\kappa}}_{\lambda(\mu}{g^{\lambda\iota}}_{\nu}{F}_{\xi)\iota}
-\frac{1}{2}{g^\kappa}_{\lambda(\mu\nu}{{F}^{\lambda}}_{\xi)}
\nonumber\\[5pt]
&&\hskip10pt...\nonumber
\end{eqnarray}  
The linear coefficient ${\gamma}^\kappa_{\mu\nu}$ is the only non-covariant one. All the others transform like tensors.
As a whole, $\bm{\gamma}^\kappa_\mu$ transforms like a connection one-form.
In fact, in (\ref{gesf}) we recognise the auto parallel equations associated to the nonlinear connection  $\bm{\gamma}^\kappa_\mu\left(x,\dot{x}\right)$ (see Appendix \ref{B}).
Therefore, we can univocally identify $\bm{\gamma}^\kappa_\mu\left(x,\dot{x}\right)$ as the connection associated the line element (\ref{tds}). 

As usual, the corresponding curvature tensor is obtained by means of the formula
\begin{equation}
\label{curvature}
\bm{r}^\kappa_{\mu\nu}=
\partial_\mu\bm{\gamma}^\kappa_\nu
-\partial_\nu\bm{\gamma}^\kappa_\mu
-\bm{\gamma}^\lambda_\mu{\textstyle\frac{\partial}{\partial\dot{x}^\lambda}}\bm{\gamma}^\kappa_\nu
+\bm{\gamma}^\lambda_\nu{\textstyle\frac{\partial}{\partial\dot{x}^\lambda}}\bm{\gamma}^\kappa_\mu.
\end{equation}
This yields the curvature $\bm{r}^\kappa_{\mu\nu}(x,\dot{x})$ again as series in $\dot{x}$
\begin{equation}
\label{nlcu}
\bm{r}^\kappa_{\mu\nu}={r^\kappa}_{\mu\nu}
+{r^\kappa}_{\xi\mu\nu}\dot{x}^{\xi}
+...+{r^\kappa}_{\xi_1...\xi_n\mu\nu}\dot{x}^{\xi_1}...\, \dot{x}^{\xi_n}+...\, .
\end{equation}
with the first few coefficients given by
\begin{eqnarray}
&&  {r^\kappa}_{\mu\nu}=
\nabla_\mu{{F}^{\kappa}}_\nu
-\nabla_\nu{{F}^{\kappa}}_\mu
-\frac{1}{2}{g^\kappa}_{\mu\lambda}{F^{\lambda}}_{\iota}{F^{\iota}}_{\nu}
+\frac{1}{2}{g^\kappa}_{\nu\lambda}{F^{\lambda}}_{\iota}{F^{\iota}}_{\mu}
\nonumber\\[5pt]
&&  {r^\kappa}_{\xi\mu\nu}={R^\kappa}_{\xi\mu\nu}
-\nabla_\mu{g^{\kappa}}_{\lambda(\nu} {{F}^{\lambda}}_{\xi)}
+\nabla_\nu{g^{\kappa}}_{\lambda(\mu} {{F}^{\lambda}}_{\xi)}+
\nonumber\\[5pt]
&&\hskip35pt
-{g^{\lambda}}_{\iota(\mu} {{F}^{\iota}}_{\xi)}
{g^{\kappa}}_{\theta(\nu} {{F}^{\theta}}_{\lambda)}
+{g^{\lambda}}_{\iota(\nu} {{F}^{\iota}}_{\xi)}
{g^{\kappa}}_{\theta(\mu} {{F}^{\theta}}_{\lambda)}
\nonumber\\[5pt]
&&\hskip35pt
-2{{F}^{\lambda}}_{\mu}{\gamma}^\kappa_{\nu\lambda\xi}
+2{{F}^{\lambda}}_{\nu} {\gamma}^\kappa_{\mu\lambda\xi}
\nonumber\\[5pt]
&&\hskip25pt...\nonumber
\end{eqnarray}
where ${R^\kappa}_{\xi\mu\nu}=\partial_\mu{\Gamma^\kappa}_{\nu\xi}-
\partial_\nu{\Gamma^\kappa}_{\mu\xi}-
{\Gamma^\lambda}_{\mu\xi}{\Gamma^\kappa}_{\nu\lambda}+
{\Gamma^\lambda}_{\nu\xi}{\Gamma^\kappa}_{\mu\lambda}$
is the Riemann tensor associated to the Christoffel-Levi-Civita connection ${\Gamma^\kappa}_{\mu\nu}$.
As in standard Riemannian geometry, it is possible to construct a contracted curvature tensor, the analogue of the Ricci tensor $R_{\mu\nu}={R^\kappa}_{\mu\kappa\nu}$, as 
\begin{equation}
\label{nRicci}
\bm{r}_{\mu}=\bm{r}^\kappa_{\kappa\mu}={r}_{\mu}
+{r}_{\xi\mu}\dot{x}^{\xi}
+...+{r}_{\xi_1...\xi_n\mu}\dot{x}^{\xi_1}...\, \dot{x}^{\xi_n}+...\, .
\end{equation} 
with
\begin{eqnarray}
&&  {r}_{\mu}=
\nabla_\kappa{{F}^{\kappa}}_\mu
-\frac{1}{2}{g^\kappa}_{\kappa\lambda}{F^{\lambda}}_{\iota}{F^{\iota}}_{\mu}
+\frac{1}{2}{g^\kappa}_{\mu\lambda}{F^{\lambda}}_{\iota}{F^{\iota}}_{\kappa}
\nonumber\\[5pt]
&&  {r}_{\xi\mu}={R}_{\xi\mu}
-\nabla_\kappa{g^{\kappa}}_{\lambda(\mu} {{F}^{\lambda}}_{\xi)}
+\nabla_\mu{g^{\kappa}}_{\lambda(\kappa} {{F}^{\lambda}}_{\xi)}+
\nonumber\\[5pt]
&&\hskip35pt
-{g^{\lambda}}_{\iota(\kappa} {{F}^{\iota}}_{\xi)}
{g^{\kappa}}_{\theta(\mu} {{F}^{\theta}}_{\lambda)}
+{g^{\lambda}}_{\iota(\mu} {{F}^{\iota}}_{\xi)}
{g^{\kappa}}_{\theta(\kappa} {{F}^{\theta}}_{\lambda)}
\nonumber\\[5pt]
&&\hskip35pt
-2{{F}^{\lambda}}_{\kappa}{\gamma}^\kappa_{\mu\lambda\xi}
+2{{F}^{\lambda}}_{\mu} {\gamma}^\kappa_{\kappa\lambda\xi}
\nonumber\\[5pt]
&&\hskip10pt...\nonumber
\nonumber
\end{eqnarray}
Correspondingly, it is also possible to define a curvature scalar by contraction with $\dot{x}^\mu$, 
\begin{equation}
\label{curvaturescalar}
\bm{r}=\bm{r}_{\mu}\dot{x}^\mu,
\end{equation} 
producing 
the symmetrisation of all components ${r}_{\xi_1...\xi_n\mu}$ of $\bm{r}_{\mu}$. This curvature scalar should not  be confused with the analogue of the scalar curvature $R=g^{\mu\nu}R_{\mu\nu}$ of standard (pseudo-)Riemannian geometry, whose generalisation is less straightforward. 

\subsection*{Quadratic restriction as quadratic approximation}
If all the fields $g_{\mu_1...\mu_k}$ except $g_{\mu\nu}$ are set equal to zero,
the geodesic equations (\ref{sge}), 
the connection (\ref{nlco}) and the curvature (\ref{nlcu}), respectively reduce to the geodesic equations, connection and curvature of standard (pseudo-)Riemannian geometry.
 Correspondingly, the line element (\ref{tds}) reduces to the integrand $\frac{1}{2}g_{\mu\nu}\dot{x}^\mu\dot{x}^\nu d\tau$ of the variational principle introduced by Levi-Civita to describe  null geodesics in general relativity \cite{Levi-Civita1929}.  In the context of standard Riemannian geometry, with a positive definite metric tensor, this is completely equivalent to Riemann's \emph{quadratic restriction}. In fact the solutions of  Levi-Civita's variational principle admit the first integral  $g_{\mu\nu}\dot{x}^\mu\dot{x}^\nu=2E$, so that 
\begin{equation}
\label{L-CvsR}
\delta\int {\textstyle \frac{1}{2}}g_{\mu\nu}\dot{x}^\mu\dot{x}^\nu d\tau =
{\textstyle \frac{1}{2}}\delta\int \sqrt{g_{\mu\nu}\dot{x}^\mu\dot{x}^\nu}  \sqrt{g_{\mu\nu}\dot{x}^\mu\dot{x}^\nu} d\tau =
 {\textstyle\sqrt{\frac{E}{2}}}\delta\int \sqrt{g_{\mu\nu}\dot{x}^\mu\dot{x}^\nu} d\tau.
\end{equation}
Conversely, when indefinite signatures are considered, Levi-Civita's variational principle becomes slightly more general. In fact, the corresponding geodesic equations coincide with the geodesic equations of standard (pseudo-)Riemannian geometry for $E\neq0$, but also include the case of null geodesics with the correct parameterisation when $E=0$ \cite{Schroedinger1956}.  The reading of Levi-Civita's integrand as a line element, is therefore an improvement of Riemann's \emph{quadratic restriction}. 

 Under this perspective, standard (pseudo-)Riemannian geometry appears as the quadratic approximation to a more general Riemannian geometry, that can be developed along the very same lines by relaxing the hypothesis of \emph{homogeneity}, but not the one of \emph{symmetry}. 
 As the quadratic term frequently provides an excellent approximation on sufficiently small scales, this suggests why Riemann's \emph{quadratic restriction} is so effective in the description of geometry.  It is effective to such an extent, that when complemented with space(time) dimension, metric signature and dynamical equations,  it is one and the same with Einstein's general theory of relativity. It is then natural to wonder whether Riemannian geometry without the hypotheses of \emph{homogeneity }and \emph{symmetry} can provide a deeper insight into the study of spacetime.

\section{Applications to spacetime}\label{spacetime}
The quadratic approximation to Riemannian geometry based on the hypotheses of \emph{differentiability}, \emph{indiscernibility} and \emph{symmetry},  thus without that of \emph{homogeneity}, naturally provides the mathematical framework of general relativity without the need for postulating the explicit form of the line element. If we further renounce the hypothesis of \emph{symmetry}, thus starting from a geometry uniquely based on \emph{differentiability} and \emph{indiscernibility}, we obtain the mathematical framework of a unified theory of the classical electromagnetic and gravitational fields. 
Unlike the classical attempts \cite{Goenner2004,Goenner2014,Tonnelat1966}, there is no need to introduce new fields, extra dimensions or new geometrical structures.
A logically simple and economical unified description is obtained simply by renouncing hypotheses within Riemannian geometry.
Unlike the classical attempts, it is directly evident how to identify the physical fields.
In the quadratic approximation a generic line element is equivalent to a one-parameter family of Riemann-Randers line elements $ds=\frac{1}{\varepsilon}g_\mu dx^\mu+\sqrt{g_{\mu\nu}dx^\mu dx^\nu}$. This
corresponds to a one-parameter family of Lagrangians for charged particles in the background of electromagnetic and a gravitational fields. The $\varepsilon$-dependent line element is not postulated for a single value of $\varepsilon$ as in Rander's proposal\footnote{See the discussion in the Introduction of \cite{Tonnelat1966}.}. It emerges by itself in as many copies as we need to describe particles with different charges. Also, in the quadratic approximation the connection (\ref{nlco}) reduces to the nonlinear connection put forward by Broda and Przanowski \cite{BrodaPrzanowski1985} for Riemann-Randers spaces.
Therefore, up to a proportionality constant, the field $g_\mu$ is naturally identified with the electromagnetic potential, as $g_{\mu\nu}$ is identified with the gravitational potential\footnote{In the original formulation of general relativity the metric tensor $g_{\mu\nu}$ was identified with the inertial-gravitational tensor \emph{potential} and the associated Christoffel symbols ${\Gamma^\kappa}_{\mu\nu}$ with the corresponding force \emph{field}, in complete analogy with the electromagnetic vector \emph{potential}  $\mathrm{A}_\mu$ and the force \emph{field} $\mathrm{F}_{\mu\nu}$. In later years, probably following the attempt to identify $g_{\mu\nu}$ and $\mathrm{F}_{\mu\nu}$ with the symmetric and antisymmetric parts of a same tensor, $g_{\mu\nu}$ has been increasingly referred to as the gravitational field. Under a different prospective, the analogy between the gauge structures of electromagnetism and Einstein's gravity have encouraged the identification of ${\Gamma^\kappa}_{\mu\nu}$ with $\mathrm{A}_\mu$ and of ${R^\kappa}_{\lambda\mu\nu}$ with $\mathrm{F}_{\mu\nu}$. The equation of motion of charged particles in a curved background leave little doubt of what the correct physical identification is.}.
Correspondingly, the parameter $\varepsilon$ is proportional to  the mass-to-charge ratio of particles, thus providing a geometrical meaning to electric charge.

As for standard Riemannian geometry and general relativity, the transition from the mathematical framework to the physical theory requires the specification of spacetime dimension, signature and dynamical equations. While the first two necessarily follow those of general relativity,  the choice of adequate dynamical equations is less constrained. The simplest and most natural generalization of Einstein field equations seems to me 
\begin{equation}
\label{fieldeq}
\bm{r}=\bm{t}
\end{equation}
with $\bm{r}$ the curvature scalar (\ref{curvaturescalar}) and $\bm{t}=t_\mu(x)\dot{x}^\mu+t_{\mu\nu}(x)\dot{x}\dot{x}^\nu+...$ an appropriate analytic function of the $x$ and of the $\dot{x}$, describing the distribution of charge and matter. As the identity of analytic functions requires the identity of all the respective coefficients in their series expansion, in the limit
where all $g_{\mu_1...\mu_k}$ except $g_\mu$ and $g_{\mu\nu}$ are negligible, 
(\ref{fieldeq}) reduces to the simultaneous Maxwell and Einstein field equations 
\begin{eqnarray}
&&  \nabla_\kappa{{F}^{\kappa}}_\mu=t_\mu,\label{Maxwell} \\
&& R_{\mu\nu}=t_{\mu\nu}.\label{Einstein}
\end{eqnarray}
I find it really remarkable that the Maxwell field equations (\ref{Maxwell}) minimally coupled to $g_{\mu\nu}$, naturally emerge from the geometry. 
On the other hand, it should also be noted that  equations  (\ref{Einstein}) do not correspond to the standard minimal coupling, because the electromagnetic stress-energy tensor  $F_{\mu\xi}{F_\nu}^\xi-\frac{1}{4}g_{\mu\nu}F_{\xi o}F^{\xi o}$  does not appear in their right-hand-side term. This is perfectly consistent with all the experiments conducted so far, where the propagation of light is studied ``along null-geodesics in  a prescribed [...] gravitational field which is a solution of Einstein vacuum equation[s] --- and not of the \emph{electro}-vacuum equation[s]'' \cite{HehlObukhov2001}. 

The standard minimal coupling can be obtained in the quadratic approximation from the straightforward generalisation of the standard Einstein-Maxwell Lagrangian density
\begin{equation}
\label{EM}
\mathcal{L}_{EM}=F_{\mu\nu}F^{\mu\nu}+R+...,
\end{equation}
where the unwritten terms for the fields ${F}_{\mu_1...\mu_k}$ for $k\geq4$, can be easily guessed from Rosen's expression for the Lagrangian density of the gravitational field  $F_{\mu\nu\xi}$ \cite{Rosen1940}.

Equivalently, the standard minimal coupling can be obtained, always in the quadratic approximation, by generalizing the geometrically more appealing Broda-Przanowski Lagrangian density \cite{BrodaPrzanowski1985} 
\begin{equation}
\label{BP}
\mathcal{L}_{BP}=g^\mu r_\mu+g^{\mu\nu}r_{\mu\nu}+g^{\mu\nu\xi}r_{\mu\nu\xi}+...
\end{equation}

In both cases, indices are raised by $g^{\mu\nu}$ and the spacetime volume element is chosen as $\sqrt{|\det g_{\mu\nu}|}d^4x$. Whatever our aesthetic preferences may be, such questions can only be solved experimentally.\\

In order to make contact with reality, it is also necessary to specify the proportionality constants between the geometrical and physical  fields and between the adimensional parameter $\varepsilon$ and the mass-to-charge ratio of physical particles.
Since the only unit of electromagnetic potential  that can be constructed in terms of classical fundamental constants is the Planck voltage
\begin{equation}
\label{PlankV}
\sqrt{\frac{c^4}{4\pi\epsilon_0G}}\simeq1.04\times10^{27}V,
\end{equation}
with $c$ the speed of light, $\epsilon_0$ the electric constant and $G$ the gravitational constant,
the only possibility is to set
 \begin{equation}
\label{g to A}
g_\mu=-\sqrt{\frac{4\pi\epsilon_0G}{c^4}}\mathrm{A}_{\mu}
\end{equation}
with $\mathrm{A}_{\mu}$  the electromagnetic vector potential.
Correspondingly, 
\begin{equation}
\label{epsilon}
\varepsilon=\frac{m}{e}\sqrt{4\pi\epsilon_0G},
\end{equation}
where $m$  and $e$ respectively indicate the mass and charge of the particle under consideration. For electrons $\varepsilon\simeq0.5\times10^{-21}$, largely justifying the validity of a quadratic approximation.\\

Beyond the quadratic approximation, the geometry imagined by Riemann in 1854  deprived of the hypotheses of \emph{homogeneity} and \emph{symmetry}, suggests the existence of infinitely more classical force fields, somehow the higher harmonics of a single fundamental classical field, the infinitesimal line element $ds$.  
These fields are generated by an identical rule from symmetric tensor potentials of increasing order in the very same way the electromagnetic and gravitational fields  $F_{\mu\nu}$ and $\Gamma_{\kappa\mu\nu}$ (or $F_{\mu\nu\xi}$) are generated by the respective potentials $g_\mu$ and $g_{\mu\nu}$.
Like electromagnetic and gravitational fields, these higher order fields  propagate in space as waves of speed $c$, interacting in an increasingly non-linear way as their order increases. Nonetheless, as the ratio of strength between successive order fields equals the one between gravity and electromagnetism, it must be expected that already the field $F_{\mu\nu\xi o}$ produces extremely weak and hard to measure local effects. On the other hand, these extremely weak effects could become relevant when integrated over the enormous spacetime distances  and could perhaps play a role in some of the many unresolved problems in astrophysics and cosmology. 
If we accept the paradigm shift that Riemann's \emph{quadratic restriction} is nothing but a quadratic approximation to a generic line element, these fields must somehow play a role in the large scale structure of spacetime. In this context, however, a non-perturbative approach to the dynamics of $ds$ might be more appropriate\footnote{It is always possible to parametrise the line element as $ds=\frac{\partial\sigma}{\partial\dot{x}^\mu}\dot{x}^\mu$, with $\sigma(x,\dot{x})$ an arbitrary analytic function of position and direction. The generalisation of Maxwell and Einstein field equations (\ref{fieldeq}) rewrites then as a single differential equation for the scalar field $\sigma$.}.

\section{Conclusion}\label{TheEnd}
For a long time the term `Riemannian geometry' has been used to indicate spherical geometry, probably a sign of how small the impact of Riemann's ideas was originally. 
With the advent of general relativity the term slowly shifted to current use and some twenty years ago Chern \cite{Chern1996} suggested that it should also include what we currently call Finsler geometry. In this paper I have used it in its broadest sense, to describe a differential geometry  uniquely based on the hypothesis that the distance of a point from itself is equal to zero. No \emph{quadratic restriction}, no \emph{homogeneity}, no \emph{symmetry}. As a matter of fact, this geometry is much more general than Riemann's original vision. 
Its relationship with standard Riemannian geometry is the same as that between a generic analytical function and its quadratic approximation at a given point. Nonetheless, the quadratic term in the series expansion of a generic line element plays such a unique role to justify the choice widely. In fact, Riemann's assumption of the infinitesimal Pythagorean distance formula reemerges as a first  approximation to \emph{almost every} geometry which is symmetric with respect to direction reversal, also trowing new light on this ancient fundamental theorem. Analogously, the associated Christoffel-Levi-Civita linear connection keeps on playing a pivotal role in the generalised connection.
For generic geometries the first approximation to the infinitesimal distance formulas turns instead into something that Riemann could hardly have foreseen. The lack of homogeneity introduces into geometry an adimensional parameter. 
Correspondingly, \emph{almost every} geometry is described in first approximation 
by a one-parameter family of Riemann-Randers metrics.  The ``square root of an [...] homogeneous function of the second degree in the quantities $dx$, in which the coefficients [$g_{\mu\nu}$] are continuous functions of the quantities $x$'', is now accompanied by a homogenous function of the first degree in the quantities $dx$ in which the coefficients $g_\mu$ are regular functions of the quantities $x$, multiplied by an arbitrary constant $\frac{1}{\varepsilon}$.  This additional term enters the first approximation to the generalised connection as the nonlinear term introduced by Broda and Przanowski.\\

If following the lines of general relativity we  apply this geometry to spacetime by identifying $g_{\mu\nu}$ with the gravitational potential, the additional term $g_\mu$ enters the geodetic and field equations exactly as an electromagnetic potential, naturally providing a unified description of the two classical fundamental interactions.
As $g_\mu$ and $g_{\mu\nu}$ respectively appears as the first and second order terms of the series expansion of the line element $ds$, the hierarchy between electromagnetic and gravitational interactions finds its origin in geometry. 
Moreover,  geometry turns out to be invariant under the simultaneous reversal of direction and of the sign of $\varepsilon$, so that also CPT invariance finds its origin in spacetime geometry. 
Correspondingly, as the different parity of subsequent terms in the series expansion of $ds$ requires that the coefficient of ${F^\kappa}_{\mu}$ changes its sign under direction reversal, while the one of ${\Gamma^\kappa}_{\mu\nu}$ does not, also the different attractive/repulsive nature of electromagnetic and gravitational interactions finds its origin in spacetime geometry. Equivalently, geometry naturally accounts for charge to be positive and negative, while mass only positive. Besides the electromagnetic and the gravitational interactions, this geometry suggests the existence of infinitely more force fields, extremely weak and so far unobserved,  that could possibly play a role in spacetime dynamics on a large scale.

 Quite remarkably, the whole kinematical setting --the number and type of potentials, their relation to fields, the hierarchy and properties of the interactions, the existence of electric charge, CPT invariance, the form of the curvature tensor that set equal to zero gives both the Maxwell and Einstein free field equations etc.-- entirely follows from the minimal assumption that the line element $ds$ is a regular function of position and direction, that the distance of a point from itself is equal to zero and, of course, from the geodesic hypothesis.     
However, as Riemann reminds us, the properties of space ``can only be deduced from experience". Therefore it remains to be seen whether these minimal hypotheses will lead to a better comprehension of the structure of spacetime on galactic and cosmological scales, so that they can be extended ``beyond the bounds of observation [...] in the direction of the immeasurably large".  On the other hand, as  ``the empirical notions on which the metric determinations of space are based [...] lose their validity in the infinitely small [...] it is [...] quite definitely conceivable that the metric relations of space in the infinitely small do not conform to the hypotheses of geometry[...] and [...] one ought to assume this as soon as it permits a [...] way of explaining phenomena". At the threshold of the quantum world a completely new idea of geometry is needed.

\subsection*{Aknowledgements}  I am indebted with J.K.\! Pachos and J.-P.\! Zendri for carefully reading the manuscript and related comments.
\appendix

\section*{Appendices}
\section{Poincar\'e-Rosen formalism for general relativity}\label{A}
Most, if not all presentations of general relativity largely emphasize the role of gravity as a geometric property of spacetime. Einstein himself, in his popular exposition of the theory, wrote  that ``in the presence of a gravitational field the geometry is not Euclidean"\cite{Einstein1923}. However, \emph{can we really measure the geometry of spacetime?} According to Poincar\`e we can not, as geometry is pure convention. In anticipating one of the classical tests of Einstein's theory, he reasoned that if we were to observe the bending of light rays, we do not have to necessarily ascribe the phenomenon to the curvature of space, as we can also ``modify the laws of optics and admit that light does not propagate strictly in a straight line"\cite{Poincare1905}.
Had he lived long enough to see the birth of general relativity, his reply to the geometrical interpretation of the theory would probably have been a slight extension of the very same formalism that Rosen developed at the end of the thirties 
in his bimetric theory of gravity \cite{Rosen1940}.
 
\paragraph{Gravity as geometry}
In the standard formalism of general relativity \emph{gravity is geometry}. The identification grounds on Einstein's equivalence principle, describing the impossibility of locally distinguishing between gravitational and inertial forces. The gravitational potential $g_{\mu\nu}(x)$ is thus identified with the spacetime metric, so that the length of the infinitesimal line segment connecting the points $x$ and $x+dx$ is expressed in terms of the gravitational potential as $ds=\sqrt{g_{\mu\nu}dx^\mu dx^\nu}$. On the one side, matter particles propagate along lines of minimal length, thus obeying the geodesic equation 
\begin{equation}
\label{geEq}
\ddot{x}^\kappa+{\Gamma^\kappa}_{\mu\nu}\dot{x}^\mu\dot{x}^\nu=0
\end{equation}
with 
$
{\Gamma^\kappa}_{\mu\nu}=\frac{1}{2}g^{\kappa\lambda}\left(\partial_\mu g_{\lambda\nu}+\partial_\nu g_{\mu\lambda}-\partial_\lambda g_{\mu\nu}\right)
$
the Christoffel symbols describing the (non-tensorial) inertial-gravitational force. 
On the other side, the geometry of spacetime is determined by the matter distribution by means of the Einstein equations
\begin{equation}
\label{EEq}
R_{\mu\nu}= \frac{8\pi G}{c^4}\left(T_{\mu\nu}-\frac{1}{2}T g_{\mu\nu}\right)
\end{equation}
where $T_{\mu\nu}$ is the energy-momentum density tensor of matter, $T=g^{\mu\nu}T_{\mu\nu}$ its trace, 
$R_{\mu\nu}={R^{\kappa}}_{\mu\kappa\nu}$ is the Ricci tensor and
$
{R^{\kappa}}_{\lambda\mu\nu}=
\partial_\mu{\Gamma^{\kappa}}_{\nu\lambda}-\partial_\nu{\Gamma^{\kappa}}_{\mu\lambda}+
{\Gamma^{\kappa}}_{\mu\xi}{\Gamma^{\xi}}_{\nu\lambda}-
{\Gamma^{\kappa}}_{\nu\xi}{\Gamma^{\xi}}_{\mu\lambda}
$
the Riemann curvature tensor. 

\paragraph{Gravity as an ambient field}
Let ${C^\kappa}_{\mu\nu}$($\neq{\Gamma^\kappa}_{\mu\nu}$) be an arbitrary connection on spacetime. Any connection, non necessarily flat and even not necessarily associated to a metric. Define the tensorial field
\begin{equation}
\label{ }
{F^\kappa}_{\mu\nu}={\Gamma^\kappa}_{\mu\nu}-{C^\kappa}_{\mu\nu}.
\end{equation}
This field is identified with the gravitational force field. It can be expressed in terms of the gravitational potential $g_{\mu\nu}$ (no longer identified with the metric) and the connection ${C^\kappa}_{\mu\nu}$ as
$
{F}_{\kappa\mu\nu}=g_{\kappa\lambda}{F^\lambda}_{\mu\nu}=\frac{1}{2}\left(\nabla_\mu g_{\lambda\nu}+\nabla_\nu g_{\mu\lambda}-\nabla_\lambda g_{\mu\nu}\right)
$,
with $\nabla_\kappa$  the covariant derivative associated to ${C^\kappa}_{\mu\nu}$.  Up to an overall sign, this is analogous to the expression of the electromagnetic field  $\mathrm{F}_{\mu\nu}$ in terms of its vector potential $\mathrm{A}_\mu$ and the connection ${C^\kappa}_{\mu\nu}$, $\mathrm{F}_{\mu\nu}=\nabla_\mu \mathrm{A}_\nu-\nabla_\nu \mathrm{A}_\mu$.

By rewriting ${\Gamma^\kappa}_{\mu\nu}$ as ${C^\kappa}_{\mu\nu}+{F^\kappa}_{\mu\nu}$ in (\ref{geEq}) we obtain
\begin{equation}
\label{grEq}
\ddot{x}^\kappa+{C^\kappa}_{\mu\nu}\dot{x}^\mu\dot{x}^\nu
+{F^\kappa}_{\mu\nu}\dot{x}^\mu\dot{x}^\nu=0.
\end{equation}
These correspond to the equations of motion of a particle in the curved background 
described by ${C^\kappa}_{\mu\nu}$, in the presence of the gravitational force field  ${F^\kappa}_{\mu\nu}$. The analogy (up to a conventional minus sign) with the equations
\begin{equation}
\label{cpcb}
\ddot{x}^\kappa+{C^\kappa}_{\mu\nu}\dot{x}^\mu\dot{x}^\nu-{\textstyle\frac{e}{mc^2}}{\mathrm{F}^\kappa}_{\mu}\dot{x}^\mu=0
\end{equation}
describing the motion of the charged particle in a curved background described by ${C^\kappa}_{\mu\nu}$under the influence of the electromagnetic field ${\mathrm{F}^\kappa}_{\mu}$, is evident. Clearly (\ref{geEq}) and (\ref{grEq}) are one and the same set of equations and the solutions do not depend on the choice of the connection ${C^\kappa}_{\mu\nu}$.

By rewriting ${\Gamma^\kappa}_{\mu\nu}$ as ${C^\kappa}_{\mu\nu}+{F^\kappa}_{\mu\nu}$ in (\ref{EEq}), we can similarly rewrite the Einstein equations as
\begin{equation}
\label{geEEq}
\nabla_\kappa{F^{\kappa}}_{\mu\nu}-
\nabla_\mu{F^{\kappa}}_{\kappa\nu}+
{F^{\kappa}}_{\kappa\xi}{F^{\xi}}_{\mu\nu}-
{F^{\kappa}}_{\mu\xi}{F^{\xi}}_{\kappa\nu}
=J_{\mu\nu}
\end{equation}
where $J_{\mu\nu}=\frac{8\pi G}{c^4}\left(T_{\mu\nu}-\frac{1}{2}T g_{\mu\nu}\right)-K_{\mu\nu}$ with $K_{\mu\nu}$ the Ricci curvature associated to the connection ${C^\kappa}_{\mu\nu}$. These correspond to the field equations for the gravitational force field ${F}_{\kappa\mu\nu}$ in the curved background described by ${C^\kappa}_{\mu\nu}$. The analogy with the Maxwell equations 
\begin{equation}
\label{Meq}
\nabla_\kappa {\mathrm{F}^{\kappa}}_\mu= J_\mu
\end{equation}
for the electromagnetic field $\mathrm{F}_{\mu\nu}$ in the curved background described by ${C^\kappa}_{\mu\nu}$, is again evident.
Clearly (\ref{EEq}) and (\ref{geEEq}) are one and the same set of equations and the solutions do not depend on the choice of ${C^\kappa}_{\mu\nu}$. 

The  impossibility of locally distinguishing between inertial and gravitational forces does not necessarily imply that these are one and the same. It just implies that inertial and gravitational forces are interchangeable.\\

In agreement with Poincar\'e, we are therefore inclined to conclude that geometry is purely conventional. 
Nonetheless, as the background geometry ${C^\kappa}_{\mu\nu}$ does not influence physical phenomena and the equations of the theory are simpler and easier to solve in the traditional formalism, we are led to prefer the convention that identifies the gravitational field with the geometry of spacetime. This was ultimately Einstein's point of view \cite{Lehmkuhl2014}, in slight contrast with Poincar\'e, who claimed that the choice of Euclidean geometry is always the most convenient. 

\section{Essentials on connections}\label{B}
The general concepts of connection and curvature are usually introduced in the rather abstract language of fiber bundles \cite{KobayashiNomizu,KolarMichorSlovak}. This makes the subject hard for non specialists to access. Since in the present discussion only local aspects matter, hereafter I will offer an essential introduction to the subject, only in terms of (local) coordinate transformations of vectors and tensors, pretty much in the spirit of classical textbooks of general relativity. 

Consider a space parameterised by $n+m$ coordinates. Divide the coordinates in two groups $(x^\mu,y^i)$ with $\mu=1,...,n$ and $i=1,...,m$, and restrict attention to the subspace $M$ parameterised by the $x$ \cite{Maraner&Pachos2008}. This subspace remains unchanged under the coordinates transformations that re-express the $x$ in terms of themselves and transform the $y$ arbitrarily
\begin{equation}
\label{ct}
\left\{
\begin{array}{l}
   x^\mu\to\bar{x}^\mu(x)\\
   y^i\to \bar{y}^i(x,y)   
\end{array}
\right.
\hskip15pt
\left\{
\begin{array}{l}
   \bar{x}^\mu\to x^\mu(\bar{x})\\
   \bar{y}^i\to y^i(\bar{x},\bar{y})   
\end{array}
\right..
\end{equation}
The Jacobian of such transformation is block-diagonal
\begin{equation}
\label{ }
\displaystyle
 {\mathbf{J}}=\left(
\begin{array}{cc}
   \frac{\partial x^\nu}{\partial \bar{x}^\mu}(\bar{x}) &  
   \frac{\partial y^j}{\partial \bar{x}^\mu}(\bar{x},\bar{y})  \\[5pt]
    0 &  
    \frac{\partial y^j}{\partial \bar{y}^i}(\bar{x},\bar{y})
\end{array}
\right)
\hskip15pt
{{\mathbf{J}^{-1}}}=\left(
\begin{array}{cc}
   \frac{\partial \bar{x}^\nu}{\partial x^\mu}(x) &  
   \frac{\partial \bar{y}^j}{\partial x^\mu}(x,y)  \\[5pt]
    0 &  
    \frac{\partial \bar{y}^j}{\partial y^i}(x,y)
\end{array}
\right),
\end{equation}
producing covariant and contravariant  $(n+m)$-vectors  to respectively transform as
\begin{equation*}
(\mathrm{v}_\mu,\mathrm{v}_i)\to \left(\frac{\partial x^\nu}{\partial \bar{x}^\mu}\mathrm{v}_\nu+
  \frac{\partial y^j}{\partial \bar{x}^\mu}\mathrm{v}_j 
    , \frac{\partial y^j}{\partial \bar{y}^i}\mathrm{v}_j\right)\hskip10pt
(\mathrm{v}^\mu,\mathrm{v}^i)\to \left(\mathrm{v}^\nu  \frac{\partial \bar{x}^\mu}{\partial x^\nu}  
  , \mathrm{v}^\mu  \frac{\partial \bar{y}^i}{\partial x^\mu}+ \mathrm{v}^j\frac{\partial \bar{y}^i}{\partial y^j}\right).
\end{equation*}
While  the last $m$ components of a  covariant $(n+m)$-vector 
transform as a covariant $m$-vector, the first $n$ components do not transform as a covariant $n$-vector.
 Also, the first $n$ components of a contravariant $(n+m)$-vector transform as a controvariant $n$-vector, while its last $m$ components do not transform as contravariant $m$-vector.

\paragraph{Connection} In order to construct covariant $n$-vectors and contravariant $m$-vectors out of $(n+m)$-vectors, it is necessary to balance the transformation rules by considering linear combinations of the form $\mathrm{v}_\mu-\bm{\gamma}_\mu^i\mathrm{v}_i$ and $\mathrm{v}^\mu \bm{\gamma}_\mu^i+\mathrm{v}^i$ with the 
$n\times m$ coefficients  $\bm{\gamma}_\mu^i(x,y)$ transforming as

\begin{equation}
\label{connection}
\bm{\gamma}_\mu^i\to 
\frac{\partial x^\nu}{\partial \bar{x}^\mu}\bm{\gamma}_\nu^j
\frac{\partial \bar{y}^i}{\partial y^j}
-\frac{\partial x^\nu}{\partial \bar{x}^\mu}\frac{\partial\bar{y}^i}{\partial x^\nu}.
\end{equation}
The choice of $n\times m$ coefficients $\bm{\gamma}_\mu^i(x,y)$ transforming in this way, corresponds to the assignment of a \emph{connection}. This assignment also allows to construct lower dimensional tensors from $(n+m)$-tensors of any order. 

\paragraph{Curvature} In terms of the connection $\bm{\gamma}_\mu^i$ it is always possible to construct a genuine lower dimensional (mixed) tensor $\bm{r}_{\mu\nu}^i(x,y)$, called the \emph{curvature} of the connection, 
\begin{equation}
\label{curvature}
\bm{r}_{\mu\nu}^i=
\partial_\mu \bm{\gamma}_\nu^i - \partial_\nu \bm{\gamma}_\mu^i
-\bm{\gamma}_\mu^j \partial_j \bm{\gamma}_\nu^i +\bm{\gamma}_\nu^j \partial_j \bm{\gamma}_\mu^i.
\end{equation} 
In fact, it is immediate to check that under the coordinate transformation (\ref{ct}) the curvature transforms as
\begin{equation}
\label{ }
\bm{r}_{\mu\nu}^i\to 
\frac{\partial x^\kappa}{\partial \bar{x}^\mu}
\frac{\partial x^\lambda}{\partial \bar{x}^\nu}
\bm{r}_{\kappa\lambda}^j\frac{\partial \bar{y}^i}{\partial y^j}.
\end{equation}

\paragraph{Riemannian geometry}
Consider now the particularly important case where  $y$ parameterize the tangent space at the point $x$ to the space $M$. In this case $m=n$ and the Latin indices $i$ coincide with the Greek indices $\mu$. In the standard basis the coordinate transformation (\ref{ct}) takes the form
\begin{equation}
\label{ctTB}
\left\{
\begin{array}{l}
   x^\mu\to\bar{x}^\mu(x)\\\displaystyle
   y^\mu\to \bar{y}^\mu=y^\nu  \frac{\partial \bar{x}^\mu}{\partial x^\nu}(x)
\end{array}
\right.
\hskip15pt
\left\{
\begin{array}{l}
   \bar{x}^\mu\to x^\mu(\bar{x})\\\displaystyle
   \bar{y}^\mu\to y^\mu=\bar{y}^\nu\frac{\partial x^\mu}{\partial \bar{x}^\nu}(x)
\end{array}
\right.
\end{equation}
and the connection transformation rule becomes
\begin{equation*}
\bm{\gamma}_\mu^\kappa\to 
\frac{\partial x^\nu}{\partial \bar{x}^\mu}
\bm{\gamma}_\nu^\lambda
\frac{\partial \bar{x}^\kappa}{\partial x^\lambda}
+
\frac{\partial^2 x^\xi}{\partial \bar{x}^\mu \partial \bar{x}^\nu}
\frac{\partial \bar{x}^\kappa}{\partial x^\xi}
\bar{y}^\nu.
\end{equation*}
Under these circumstances the connection $\bm{\gamma}_\mu^\kappa(x,y)$ 
defines an infinitesimal parallel transport of the vector $y$ in the direction parameterised by $x^\mu$ as $\partial_\mu y^\kappa+\bm{\gamma}_\mu^\kappa(x,y)$. The lines $x(\tau)$ such that their tangent vector $\dot{x}(\tau)$ is parallely transported along them at each point are then called auto parallel lines. They satisfy the \emph{auto parallel equations}
\begin{equation}
\label{apeq}
\ddot{x}^\kappa+\bm{\gamma}^\kappa_\mu\dot{x}^\mu=0.
\end{equation}

Standard Riemannian  geometry corresponds to the choice of a connection linear in the $y$, 
\begin{equation}
\label{Rconnection}
\bm{\gamma}_\mu^\kappa(x,y)={\Gamma^\kappa}_{\mu\nu}(x)y^\nu
\end{equation}
with the coefficients ${\Gamma^\kappa}_{\mu\nu}={\Gamma^\kappa}_{\nu\mu}$ symmetric in the two lower indices. The connection transformation rule becomes then the  standard transformation rule of the Christoffel symbols
\begin{equation*}
{\Gamma^\kappa}_{\mu\nu}y^\nu\to
\left(\frac{\partial x^\rho}{\partial \bar{x}^\mu}
\frac{\partial x^\sigma}{\partial \bar{x}^\nu}
{\Gamma^\lambda}_{\rho\sigma} 
\frac{\partial \bar{x}^\kappa}{\partial x^\lambda} 
+\frac{\partial^2 x^\xi}{\partial \bar{x}^\mu\partial \bar{x}^\nu}
     \frac{\partial \bar{x}^\kappa}{\partial x^\xi}\right)\bar{y}^\nu,
\end{equation*}
while the curvature (\ref{curvature}) reduces to the Riemann tensor
\begin{equation}
\label{Rcurvature}
\bm{r}_{\mu\nu}^\kappa=\left(\partial_\mu{\Gamma^\kappa}_{\nu\xi}-
\partial_\nu{\Gamma^\kappa}_{\mu\xi}-
{\Gamma^\lambda}_{\mu\xi}{\Gamma^\kappa}_{\nu\lambda}+
{\Gamma^\lambda}_{\nu\xi}{\Gamma^\kappa}_{\mu\lambda}\right)y^\xi
={R^\kappa}_{\lambda\mu\nu}y^\lambda.
\end{equation}
The auto parallel equations (\ref{apeq}) coincides with the geodesic  equations 
(\ref{geEq}).

More general choices of $\bm{\gamma}_\mu^\kappa$, possibly non linear in the $y$, produce more general geometries.



\begin{thebibliography}{99}
\bibitem{Jammer1954}
Jammer, M. Concepts of space. Dover, New York, 1993.
\bibitem{Riemann}
Riemann, B. Ueber die Hypothesen, welche der Geometrie zu Grunde liegen, \emph{Abhandlungen der K\"{o}niglichen Gesellschaft der Wissenschaften zu G\"{o}ttingen}, \textbf{13} (1868), 133-150. 
Reprinted in the English translation by M. Spivack in \cite{Pesic2007}.
\bibitem{Pesic2007}
Pesic, P. Beyond geometry: Classic papers from Riemann to Einstein. Courier Corporation, 2007.
\bibitem{Rund2012}
Rund, H. The differential geometry of Finsler spaces. Vol. 101. Springer Science \& Business Media, 2012.
\bibitem{BaoChernShen2012}
Bao, D., Chern, S-S., Shen, Z. An introduction to Riemann-Finsler geometry. Vol. 200. Springer Science \& Business Media, 2012.
\bibitem{Randers1941}
Randers, G. On an asymmetrical metric in the four-space of general relativity. \emph{Physical Review} \textbf{59} (1941), 19-199.
\bibitem{Goenner2004}
Goenner, H.F.M. On the history of unified field theories. \emph{Living reviews in relativity} \textbf{7} (2004): 2.
\bibitem{Goenner2014}
Goenner, H.F.M. On the history of unified field theories. Part II. (ca. 1930-ca. 1965). \emph{Living Reviews in Relativity} \textbf{17} (2014): 5.
\bibitem{Tonnelat1966}
Tonnelat, M.A. Einstein's Theory of Unified Fields. Routledge, 2014.
\bibitem{Einstein1921}
Einstein A.  Geometrie und Erfahrung. In: Geometrie und Erfahrung. Springer, Berlin, Heidelberg, 1921. Reprinted in the English translation by S. Bargmann in \cite{Pesic2007}.
\bibitem{Minkowski1908}
Minkowski, H. \emph{Raum und Zeit}. Physikalishe Zeitrift \textbf{10} (1909) 104-111.
Reprinted in English translation by W. Perrett and G. B. Jeffrey in The principle of relativity. Courier Corporation, 2013.
\bibitem{Chern1996}
Chern, S.-S. Finsler geometry is just Riemannian geometry without the quadratic restriction. \emph{Notices of the American Mathematical Society} \textbf{43} (1996): 959-963.
\bibitem{Rutz1993}
Rutz, S.F. A Finsler generalisation of Einstein's vacuum field equations. \emph{General Relativity and  Gravitation} \textbf{25} (1993): 1139-1158.
\bibitem{Asanov2012}
Asanov, G. S. Finsler geometry, relativity and gauge theories. Vol. 12. Springer Science \& Business Media, 2012.
\bibitem{Pfeifer&Wohlfarth2012}
Pfeifer, C., Wohlfarth M.N.R. Finsler geometric extension of Einstein gravity. \emph{Physical Review D} \textbf{85} (2012): 064009.
\bibitem{Stephenson1957}
Stephenson, G. La g\'eometrie de Finsler et les th\'eories du champs unifi\'e. \emph{Ann. Inst. Henri Poincar\'e} \textbf{15} (1957) 205-215.
\bibitem{Levi-Civita1929}
Levi-Civita, T. The Absolute Differential Calculus. Blackie and Son, 1929.
\bibitem{Schroedinger1956}
Schr\"{o}dinger, E. Expanding universes. Cambridge University Press, 1956.
\bibitem{BrodaPrzanowski1985}
Broda, B., Przanowski M. Electromagnetic field as a nonlinear connection. \emph{Acta Phys.\ Polon.\ } \textbf{17} (1985) 481-484.
\bibitem{HehlObukhov2001}
Hehl, F.W., Obukhov, Y.N. How does the electromagnetic field couple to gravity, in particular to metric, nonmetricity, torsion, and curvature?. in Gyros, Clocks, Interferometers ...: Testing Relativistic Graviy in Space,
L\"ammerzahl C., Everitt, C.W.F., Hehl, F.W., eds.
Springer, 2001, 479-504.
\bibitem{Rosen1940}
Rosen, N. General relativity and flat space. I \& II \emph{Physical Review} \textbf{57} (1940)147-153.
\bibitem{Einstein1923}
Einstein, A. The meaning of relativity. Princeton University Press, Princeton 1923.
\bibitem{Poincare1905}
Poincar\'e, H. Science and hypothesis. The Walter Scott Publishing Co., New York 1905.
\bibitem{Lehmkuhl2014}
Lehmkuhl, D. Why Einstein did not believe that general relativity geometrizes gravity. \emph{Studies in History and Philosophy of Science Part B: Studies in History and Philosophy of Modern Physics} \textbf{46} (2014) 316-326.
\bibitem{KobayashiNomizu}
Kobayashi, S., Nomizu, K. Foundations of differential geometry, Vol. 1. 
Wiley-Interscience, 1996
\bibitem{KolarMichorSlovak}
Kolar, I., Michor, P., and Slovak, J. Natural operations in Differential Geometry. Springer-Verlag, 1993.
\bibitem{Maraner&Pachos2008}
Maraner, P., Pachos, J.K. Universal features of dimensional reduction schemes from general covariance breaking. \emph{Annals of Physics} \textbf{323} (2008) 2044-2072.
\end{thebibliography}
\end{document}